\documentclass[a4paper,12pt]{article}

\usepackage{graphicx}
\usepackage{amsmath}
\usepackage{geometry}
\usepackage{fancyhdr}
\usepackage{setspace}
\usepackage{titlesec}  % For title formatting
\usepackage[pagebackref,breaklinks,colorlinks]{hyperref}
\usepackage{booktabs} 
\usepackage[capitalize]{cleveref}
\crefname{section}{Sec.}{Secs.}
\Crefname{section}{Section}{Sections}
\Crefname{table}{Table}{Tables}
\crefname{table}{Tab.}{Tabs.}

\usepackage{tocloft} % Package for customizing ToC
\usepackage{cite} % For citations, if needed
\titleformat{\section}{\normalfont\Large\bfseries}{\thesection}{1em}{}

\title{
    \vspace{2cm} % Adjust vertical space
    \vspace{1cm} % Adjust vertical space after the logo
    \textbf{\Huge An Overview of Machine Learning-Driven Resource Allocation in IoT Networks} \\
    \vspace{1cm} % Adjust vertical space
    \large EESM5650 - Digital Communication Networks and Systems \\
    \vspace{0.5cm} % Adjust vertical space
}
\author{Zhengdong Li \\ Department of Electronic and Computer Engineering, HKUST\\ zlifd@connect.ust.hk}
\date{}

\begin{document}

% Title Page
\maketitle
\thispagestyle{empty}
\newpage

% Table of Contents
% Start page numbering from the Table of Contents
\setcounter{page}{1}  % Start counting from 1
\tableofcontents
\newpage

% 1. Abstract
\section{Abstract}
%-----------------------------------------------------------------------
    In the wake of disruptive IoT technologies generating massive amounts of diverse data, Machine Learning (ML) will play a crucial role in bringing intelligence to Internet of Things (IoT) networks. This paper provides a comprehensive analysis of the current state of resource allocation within IoT networks, focusing specifically on two key categories: Low-Power IoT Networks and Mobile IoT Networks. We delve into the resource allocation strategies that are crucial for optimizing network performance and energy efficiency in these environments. Furthermore, the paper explores the transformative role of Machine Learning (ML), Deep Learning (DL), and Reinforcement Learning (RL) in enhancing IoT functionalities. We highlight a range of applications and use cases where these advanced technologies can significantly improve decision-making and optimization processes. In addition to the opportunities presented by ML, DL, and RL, we also address the potential challenges that organizations may face when implementing these technologies in IoT settings. These challenges include crucial accuracy, low flexibility and adaptability, and high computational cost, etc. Finally, the paper identifies promising avenues for future research, emphasizing the need for innovative solutions to overcome existing hurdles and improve the integration of ML, DL, and RL into IoT networks. By providing this holistic perspective, we aim to contribute to the ongoing discourse on resource allocation strategies and the application of intelligent technologies in the IoT landscape.

% 2. Background, Motivation and History
\section{Background and Motivation}
%-----------------------------------------------------------------------
    Internet of Things, or IoT, refers to the connection of massive smart devices to internet with intensive heterogeneous networks. According to recent estimation \cite{hussain2020machine}, the world is currently home to approximately 25 billion embedded smart systems, generating 50 trillion GB of data within a global population of 4 billion connected people. This remarkable proliferation of connected devices and data is driven by the rapid advancement of the IoT, particularly in the wireless domain. It has the potential to profoundly shape the future of our smart world, helping to connect and transform various aspects of our daily lives and businesses. From the ubiquity of smartphones to the increasing integration of IoT in housing and security systems, the IoT is revolutionizing how we interact with and leverage technology. Analysts \cite{manyika20152025} estimate that the IoT will generate between \$3.9 trillion and \$11.1 trillion in output by the year 2025, as the technology becomes increasingly pervasive across diverse environments, such as retail, smart cities, and industrial settings \cite{ullah2024integration}. This exponential growth is further underscored by the projected increase in the number of IoT devices, which is expected to surpass 750 billion globally. In fact, the rate of IoT device adoption is so rapid that it is estimated that 127 new IoT devices are added in every second since 2020 \cite{manyika20152025}. As the IoT ecosystem continues to expand and evolve, it is clear that massive scale IoT will play a crucial role in shaping the smart world of the future, revolutionizing the way we live, work, and interact with our surroundings. 

    As the scale of IoT deployments continues to grow, it has become increasingly critical to address the challenges that arise from this massive increment of connected devices. Network congestion, storage architecture constraints, and the need for efficient data communication protocols have emerged as pressing concerns \cite{bandyopadhyay2011internet} \cite{shanthamallu2017brief}. Conventional approaches to resource management and access control may prove inadequate in the face of the sheer volume and diversity of IoT devices. Furthermore, some IoT applications, such as intelligent transportation systems and remote surgery demand low-latency communication and ultra-reliable connectivity, necessitating the development of innovative resource allocation techniques that can ensure the reliable and timely delivery of data. Thus, the demand of real time processing and massive channels access contribute the key factors for the motivation of this survey paper. In this context, the exploration of machine learning-driven resource allocation strategies in IoT networks holds immense promise, offering the potential to optimize network performance, enhance user experience, and enable the realization of a truly smart and connected world.

\section{Current Resource Allocation in IoT Networks}
%-------------------------------------------------------------------
    Journal Paper \cite{hussain2020machine} is the main reference for this project. In this section, we will discuss the existing methods for resource allocation in IoT network, including Low-Power IoT Network and Mobile IoT Network. 

\subsection{Low-Power IoT Network}

    While IoT networks offer both long-range and short-range communication capabilities through various technologies, there are many applications that require the energy efficiency of short-range technologies while maintaining long-range communication. As a result, low-power IoT networks supporting long-range communications have become essential, such as Low-Power Wide Area Networks (LPWANs). LPWANs provide long-range communication by limiting data rates and energy usage. It has been studied widely for the purpose of diverse smart applications.

    Several technologies in both licensed, such as LTE-M and unlicensed, such as Sigfox and LoRa bands are being considered for low-power networks,  dealing with different resource allocation challenges individually. For instance, \cite{raza2017low} has examined the emerging LPWA technologies and the standardization efforts by various organizations like IEEE and industry group like LoRa Alliance. These networks typically have a large number of nodes, and collisions are mitigated through scheduling and duty cycling \cite{mekki2019comparative}. Their study evaluated the leading LPWAN technologies like Sigfox and LoRa, which offer advantages in terms of capacity, battery life and cost.

\subsection{Mobile IoT Network}

    Mobile IoT is built on top of traditional IoT with the extension of mobility \cite{alnahdi2017mobile}, where IoT services and applications can be physically transported from one place to another, resulting in higher mobility. While traditional IoT enables communication within static objects to realize various services and increase connectivity, Mobile IoT involves communication entities that are mobile and maintain their accessibility and interconnection as they move. Mobile objects, including but not limited to vehicles and robots help to add the  dimension of mobility to traditional IoT which only has static connectivity. 
    
    Common Mobile IoT applications, including smart industry with mobile robots and smart transportation with connected vehicles, has high node mobility. This introduces additional control information exchange requirements compared to traditional IoT networks, making resource allocation more challenging.

\section{Applying Machine Learning in IoT Networks}
%-----------------------------------------------------------------------
    In this section, it will first briefly introduce ML and DL and how they can be applied in IoT networks. 
    
\subsection{Introduction of ML, DL and RL}

    Machine learning (ML) is a field of Artificial Intelligence (AI) that enables computers to learn and improve from experience without being explicitly programmed. There are three main categories of ML techniques, naming Supervised Learning, Unsupervised Learning and Reinforcement Learning respectively. 

    In Supervised Learning, the algorithm is provided with labeled training data, which includes both the input features and the desired output or target variable. The algorithm learns a mapping function from the input to the output, and can then make predictions on new, unseen data. Example includes image classification. 

    In Unsupervised Learning, the algorithms are given unlabeled data, without any predetermined output variables. The goal is to discover hidden patterns, structures, or groupings within the data. Common unsupervised learning techniques include dimensionality reduction, like feature extraction.

    In Reinforcement Learning (RL), it is a type of ML where an agent learns by interacting with an environment and receiving feedback in the form of rewards or penalties. The agent learns to take actions that maximize the cumulative reward or minimize the penalties over time. It is often used in sequential decision-making problems, such as AlphaGo, a computer Go program developed by DeepMind, played a historic five-game match against world-class Go player Lee Sedol in 2016 by wining the match 4-1 and demonstrating the remarkable capabilities of AI in the ancient game of Go \cite{sormani2023interfacing}. 

\begin{figure}
    \centering
    \includegraphics[width=1\linewidth]{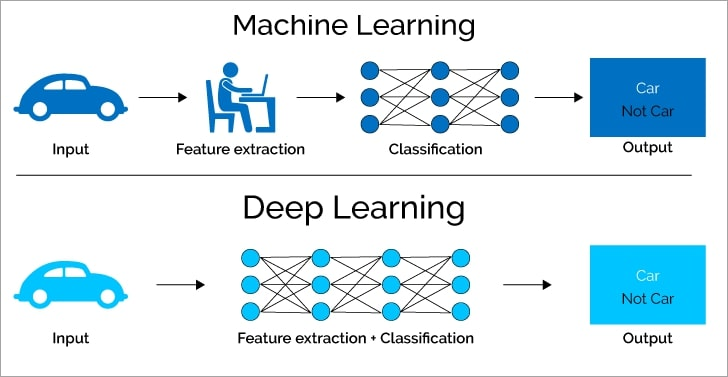}
    \caption{A graph representation of ML vs DL.}
    \label{fig:ML_DL}
\end{figure}

    Deep learning (DL) is a branch of machine learning, as shown in Fig. \ref{fig:ML_DL} \cite{levity2024}, that uses artificial neural networks with multiple hidden layers to learn and make predictions from data. It has been widely used in the fields of computer vision, natural language processing and speech recognition, etc. 

\subsection{Application of ML, DL and RL in IoT network}

    DL has been widely applied to advanced wireless technologies like Multi-Input Multi-Output (MIMO) and Non-Orthogonal Multiple Access (NOMA).  DL-based channel estimation and code construction techniques have been proposed to improve performance. For example, \cite{huang2018deep} used DL in MIMO for high resolution channel prediction. \cite{kim2018deep} used DL as helps to develop Sparse Code Multiple Access (SCMA) for code-based NOMA, which constructs an adaptive code book to decrease the Bit Error Rate (BER) of the NOMA system. It has developed and formulated an effective decoding strategy using a deep neural networks-based encoder and decoder. Their  simulation results demonstrate a significantly lower BER and lower computational time.

    ML is a powerful tool for solving resource allocation problems in IoT networks. It involve a large amount of data collected from various connected devices, which can be leveraged to train ML algorithms to optimize resource allocation.

    For example, \cite{wang2018machine} used cloud computing and ML techniques to perform resource allocation in the wireless network, applying the  proposed methods for beam allocation in a wireless system. \cite{bohez2015discrete} used multiple ML algorithms in a cloud-based network to optimize the networking from the perspectives of distributing computing and processing functions across the network entities. Besides, different ML-based resource management algorithms have been developed for the video application in IoT network. \cite{testolin2014machine} presented a combination of both supervised and unsupervised learning algorithms for the optimization on the Quality-of-Experience (QoE) for end-users while video streams are transmitted over a wireless channel. 
    
    Furthermore, \cite{ahmed2019deep} introduced a centralized deep RL-based power allocation scheme for a multi-cell system. While maximizing the overall network throughput, they particularly employed a deep Q-Learning approach to attain near-optimal power allocation. \cite{ye2019deep} introduced a multi-agent deep RL for resource allocation in vehicle-to-vehicle communications. To establish the mapping between each vehicle's local observations, like Channel State Information (CSI) and interference levels which will influence the resource allocation decisions. Deep RL was employed such that every vehicle, or namely the agent, makes optimal transmission power decisions by interacting with its environment. And Tab. \ref{tab:ML_DL_table} shows the summary of the ML, DL and RL examples in IoT network.

    \begin{table}
        \centering
        \scriptsize
        \begin{tabular}{ccc}
             \toprule
             \textbf{ML/DL/RL?}& \textbf{References} & \textbf{Key Contributions}\\
             \midrule
             DL & \cite{huang2018deep} & to perform high resolution channel prediction in MIMO\\
             DL & \cite{kim2018deep} & to develop SCMA for code-based NOMA \\
             ML & \cite{wang2018machine} & to perform resource allocation in the wireless network\\
             ML & \cite{bohez2015discrete} & to optimize networking of distributing computing and processing functions \\
             ML & \cite{testolin2014machine} & to optimize QoE \\
             RL & \cite{ahmed2019deep} & to maximize overall network throughput by Q-Learning\\
             RL & \cite{ye2019deep} & to makes optimal transmission power decisions via multi-agents\\
             \bottomrule
        \end{tabular}
        \caption{Summary of various ML and DL examples in IoT network}
        \label{tab:ML_DL_table}
    \end{table}

\section{Potential Challenges and Future Research Tread}
%-----------------------------------------------------------------------

    In this section, we will further discuss some potential challenges for ML and DL in IoT networks and possible future research tread in this area.

    Firstly, the accuracy of ML and DL models is a crucial factor, especially in applications where precision is paramount, such as medical surgery. In these critical domains, false positives or inaccurate predictions can have devastating consequences, potentially putting lives at risk. The training process for such models must be meticulously designed to ensure the model can reliably extract the right features from the data and learn effectively. This is a vulnerable factor that requires extensive testing and validation to ensure the reliability of the ML models and mitigate the risks associated with inaccurate predictions before applying them in IoT network.

    Secondly, the highly specialized nature of ML and DL models poses a challenge in IoT networks. These models are typically trained for a specific task and may not be easily applicable to multiple applications. Restructuring the entire architecture of the model to accommodate different tasks can be time-consuming and expensive, which may not be feasible in resource-constrained IoT environments. This limitation can hinder the flexibility and adaptability of DL models in IoT networks.

    Lastly, the intensive data requirements and high computational resources needed for training large-scale DL and ML models can be a significant challenge in IoT networks. The cost of the necessary GPU resources may be prohibitive, especially for smaller-scale IoT deployments. This financial burden can limit the adoption and deployment of advanced ML and DL techniques in IoT applications, where cost-effectiveness is a critical factor.
    
    The groundbreaking work of AI pioneers like John Hopfield and Geoffrey Hinton has revolutionized ML. They won The Nobel Prizes in Physics 2024, where the role of physics in developing the foundations of artificial neural networks has been acknowledged \cite{holme2024education}. And the transformative power of AI is also poised to reshape resource allocation within the rapidly expanding IoT networks. With billions of connected devices generating unprecedented amounts of data, the need for efficient and intelligent management of these network resources has never been more pressing. Future research will focus on leveraging advanced AI techniques, particularly deep learning (DL), to facilitate dynamic allocation of bandwidth, computing power, and energy resources across diverse IoT ecosystems. This includes developing algorithms that can learn from network patterns and user behaviors, enabling real-time optimization of resource distribution to meet the evolving demands of connected devices.

    Additionally, the integration of AI-driven resource allocation with emerging concepts like edge computing and 6G networks will be critical to address the unique challenges of the IoT paradigm. For example, in 6G networks, the usage of AI and communication help to address distributed computing, autonomous driving \cite{chen2020deep} and auto collaboration in medical assistance \cite{mucchi20206g}, etc, distributing intensive computational tasks across multiple devices and networks \cite{new2024tutorial}.

% 5. Conclusions
\section{Conclusions}
%-----------------------------------------------------------------------

    In conclusion, this paper underscores the pivotal role that ML, DL and RL play in advancing the capabilities of IoT networks. As disruptive IoT technologies continue to generate vast amounts of diverse data, effective resource allocation becomes essential for optimizing both network performance and energy efficiency, particularly within Low-Power and Mobile IoT Networks. We have explored various resource allocation strategies that leverage these advanced technologies, highlighting their transformative potential across multiple applications and use cases. However, it is also essential to acknowledge the challenges that may encounter, including issues related to accuracy, flexibility, adaptability, and computational costs. Furthermore, this paper emphasizes the importance of innovative research to address these challenges and enhance the integration of ML, DL, and RL into IoT networks. By fostering a deeper understanding of these technologies and their implications for resource allocation, we aim to contribute meaningfully to the ongoing discourse in this dynamic field, paving the way for smarter, more efficient IoT solutions in the future.

\newpage
\section{References}
% Bibliography (if required)
\bibliographystyle{plain}
\bibliography{references}  % Add a .bib file if you have references

\begin{thebibliography}{10}

\bibitem{ahmed2019deep}
Kazi~Ishfaq Ahmed and Ekram Hossain.
\newblock A deep q-learning method for downlink power allocation in multi-cell networks.
\newblock {\em arXiv preprint arXiv:1904.13032}, 2019.

\bibitem{alnahdi2017mobile}
Amany Alnahdi and Shih-Hsi Liu.
\newblock Mobile internet of things (miot) and its applications for smart environments: A positional overview.
\newblock In {\em 2017 IEEE International Congress on Internet of Things (ICIOT)}, pages 151--154. IEEE, 2017.

\bibitem{bandyopadhyay2011internet}
Debasis Bandyopadhyay and Jaydip Sen.
\newblock Internet of things: Applications and challenges in technology and standardization.
\newblock {\em Wireless personal communications}, 58:49--69, 2011.

\bibitem{bohez2015discrete}
Steven Bohez, Tim Verbelen, Pieter Simoens, and Bart Dhoedt.
\newblock Discrete-event simulation for efficient and stable resource allocation in collaborative mobile cloudlets.
\newblock {\em Simulation Modelling Practice and Theory}, 50:109--129, 2015.

\bibitem{chen2020deep}
Xiaosha Chen, Supeng Leng, Jianhua He, and Longyu Zhou.
\newblock Deep-learning-based intelligent intervehicle distance control for 6g-enabled cooperative autonomous driving.
\newblock {\em IEEE Internet of Things Journal}, 8(20):15180--15190, 2020.

\bibitem{holme2024education}
Thomas Holme.
\newblock Education implications of artificial intelligence-based chemistry and physics nobel prizes, 2024.

\bibitem{huang2018deep}
Hongji Huang, Jie Yang, Hao Huang, Yiwei Song, and Guan Gui.
\newblock Deep learning for super-resolution channel estimation and doa estimation based massive mimo system.
\newblock {\em IEEE Transactions on Vehicular Technology}, 67(9):8549--8560, 2018.

\bibitem{hussain2020machine}
Fatima Hussain, Syed~Ali Hassan, Rasheed Hussain, and Ekram Hossain.
\newblock Machine learning for resource management in cellular and iot networks: Potentials, current solutions, and open challenges.
\newblock {\em IEEE communications surveys \& tutorials}, 22(2):1251--1275, 2020.

\bibitem{kim2018deep}
Minhoe Kim, Nam-I Kim, Woongsup Lee, and Dong-Ho Cho.
\newblock Deep learning-aided scma.
\newblock {\em IEEE Communications Letters}, 22(4):720--723, 2018.

\bibitem{levity2024}
Levity.
\newblock Deep learning vs. machine learning – what’s the difference?
\newblock 2024.

\bibitem{manyika20152025}
James Manyika and Michael Chui.
\newblock By 2025, internet of things applications could have \$11 trillion impact.
\newblock {\em Insight Publications}, 2015.

\bibitem{mekki2019comparative}
Kais Mekki, Eddy Bajic, Frederic Chaxel, and Fernand Meyer.
\newblock A comparative study of lpwan technologies for large-scale iot deployment.
\newblock {\em ICT express}, 5(1):1--7, 2019.

\bibitem{mucchi20206g}
Lorenzo Mucchi, Sara Jayousi, Stefano Caputo, Elisabetta Paoletti, Paolo Zoppi, Simona Geli, and Pietro Dioniso.
\newblock How 6g technology can change the future wireless healthcare.
\newblock In {\em 2020 2nd 6G wireless summit (6G SUMMIT)}, pages 1--6. IEEE, 2020.

\bibitem{new2024tutorial}
Wee~Kiat New, Kai-Kit Wong, Hao Xu, Chao Wang, Farshad~Rostami Ghadi, Jichen Zhang, Junhui Rao, Ross Murch, Pablo Ram{\'\i}rez-Espinosa, David Morales-Jimenez, et~al.
\newblock A tutorial on fluid antenna system for 6g networks: Encompassing communication theory, optimization methods and hardware designs.
\newblock {\em IEEE Communications Surveys \& Tutorials}, 2024.

\bibitem{raza2017low}
Usman Raza, Parag Kulkarni, and Mahesh Sooriyabandara.
\newblock Low power wide area networks: An overview.
\newblock {\em ieee communications surveys \& tutorials}, 19(2):855--873, 2017.

\bibitem{shanthamallu2017brief}
Uday~Shankar Shanthamallu, Andreas Spanias, Cihan Tepedelenlioglu, and Mike Stanley.
\newblock A brief survey of machine learning methods and their sensor and iot applications.
\newblock In {\em 2017 8th International Conference on Information, Intelligence, Systems \& Applications (IISA)}, pages 1--8. Ieee, 2017.

\bibitem{sormani2023interfacing}
Philippe Sormani.
\newblock Interfacing alphago: Embodied play, object agency, and algorithmic drama.
\newblock {\em Social Studies of Science}, 53(5):686--711, 2023.

\bibitem{testolin2014machine}
Alberto Testolin, Marco Zanforlin, Michele De~Filippo De~Grazia, Daniele Munaretto, Andrea Zanella, Marco Zorzi, and Michele Zorzi.
\newblock A machine learning approach to qoe-based video admission control and resource allocation in wireless systems.
\newblock In {\em 2014 13th Annual Mediterranean Ad Hoc Networking Workshop (MED-HOC-NET)}, pages 31--38. IEEE, 2014.

\bibitem{ullah2024integration}
Inam Ullah, Deepak Adhikari, Xin Su, Francesco Palmieri, Celimuge Wu, and Chang Choi.
\newblock Integration of data science with the intelligent iot (iiot): current challenges and future perspectives.
\newblock {\em Digital Communications and Networks}, 2024.

\bibitem{wang2018machine}
Jun-Bo Wang, Junyuan Wang, Yongpeng Wu, Jin-Yuan Wang, Huiling Zhu, Min Lin, and Jiangzhou Wang.
\newblock A machine learning framework for resource allocation assisted by cloud computing.
\newblock {\em IEEE Network}, 32(2):144--151, 2018.

\bibitem{ye2019deep}
Hao Ye, Geoffrey~Ye Li, and Biing-Hwang~Fred Juang.
\newblock Deep reinforcement learning based resource allocation for v2v communications.
\newblock {\em IEEE Transactions on Vehicular Technology}, 68(4):3163--3173, 2019.

\end{thebibliography}

\end{document}